# James Webb Space Telescope segment phasing using differential optical transfer functions

Johanan L. Codona
Nathan Doble

# James Webb Space Telescope segment phasing using differential optical transfer functions

**Johanan L. Codona**[a,*] **and Nathan Doble**[b]
[a]University of Arizona, Steward Observatory, Tucson, Arizona 85721, United States
[b]Ohio State University, College of Optometry, Columbus, Ohio 43210, United States

**Abstract.** Differential optical transfer function (dOTF) is an image-based, noniterative wavefront sensing method that uses two star images with a single small change in the pupil. We describe two possible methods for introducing the required pupil modification to the James Webb Space Telescope, one using a small ($<\lambda/4$) displacement of a single segment's actuator and another that uses small misalignments of the NIRCam's filter wheel. While both methods should work with NIRCam, the actuator method will allow both MIRI and NIRISS to be used for segment phasing, which is a new functionality. Since the actuator method requires only small displacements, it should provide a fast and safe phasing alternative that reduces the mission risk and can be performed frequently for alignment monitoring and maintenance. Since a single actuator modification can be seen by all three cameras, it should be possible to calibrate the non-common-path aberrations between them. Large segment discontinuities can be measured using dOTFs in two filter bands. Using two images of a star field, aberrations along multiple lines of sight through the telescope can be measured simultaneously. Also, since dOTF gives the pupil field amplitude as well as the phase, it could provide a first approximation or constraint to the planned iterative phase retrieval algorithms.  

## 1 Introduction

The James Webb Space Telescope (JWST) (Refs. 1 and 2) primary mirror consists of 18 hexagonal mirror segments mounted via a hexapod to a rigid backplane structure. Each hexapod allows its segment to be controlled in six degrees of freedom: transverse position, piston, tip, tilt, and clocking. A seventh actuator allows the segment's radius of curvature to be adjusted. For the JWST to perform with optimal imaging resolution and sensitivity, the segments must be initially aligned and co-phased, followed by periodic alignment monitoring and updates. NIRCam[3,4] is the only instrument involved in this fine wavefront sensing and segment co-phasing procedure and, as such, is critical to the success of the mission. The fine segment phasing involves using a set of defocused star images with an iterative phase retrieval algorithm to find the segment offsets and deviations from the desired figure.[5] The alignment and phasing procedure is expected to be reliable and robust and maintain alignment to a wavefront error (WFE) of <150 nm rms (Ref. 4) to ensure diffraction-limited images. Maintaining this WFE requires wavefront sensing precision to a much higher accuracy, perhaps 10 times better or 15 nm. Although precision optics and preflight measurements should allow accurate cross-camera calibration, using NIRCam for the wavefront sensor will naturally calibrate-in any aberrations between NIRCam and the primary. The other imaging cameras, MIRI (Refs. 6 and 7) and near-infrared imager and slitless spectrograph (NIRISS),[8] may develop different non-common-path aberrations than those measured preflight, which will then be difficult to ascertain. A technique has been developed to use NIRISS for segment phasing,[9] which will somewhat relieve the critical role of NIRCam.

The method of differential optical transfer functions (dOTF)[10,11] is a technique for measuring the complex amplitude of a pupil field, including all the transmission and aberration effects between the pupil and the imaging camera. The method works by collecting two star images with a localized modification introduced into the telescope's pupil for one of the star images. The resulting images are Fourier transformed and subtracted from each other, resulting in an estimate of the complex field over the majority of the pupil, along with a second complex conjugated field image reflected about the location of the pupil change. Since the point of reflection is inside but near the edge of the pupil, a small portion of the pupil is obscured. If the modified pupil area is small, the complex field image will be crisp, but have a low signal. A larger modification area gives the pupil field cross-correlated with the field difference over the changed area, which may still be sufficient as an answer, or may require some further deconvolutional postprocessing. The pupil modification may involve transmission or phase or a combination, each giving similar results. It is of practical note that the dOTF method is insensitive to the ability to introduce repeatable, or even known, changes into the pupil. The method will be very fast to perform and can even simultaneously measure the aberrations and vignetting corresponding to many field points using only two images using a field of stars.[12,13] In this paper, we consider the possibility of using dOTF as an alternative or supplementary segment phasing technique for the JWST using any of NIRcam, MIRI, or NIRISS. Each camera can make dOTF measurements by moving a single segment actuator by <1/4 of the given camera's filter's wavelength. In the case of NIRcam, another option is possible by using the filter wheel to obscure the edge of one of the primary segments. In each case, very small modifications can give highly accurate wavefront measurements, making dOTF a simple and safe technique for phasing and maintaining the alignment of the primary segments. For the JWST, it is unlikely that there will be sufficiently serious failures with NIRCam and the secondary mirror that both the planned and contingency phase retrieval methods will be unavailable. Therefore, having the ability to use dOTF is likely to be operationally redundant. However, dOTF's availability will provide some level of mission risk reduction and can be used as a double-check on phase retrieval results as needed. Because of the expected quality of the phase retrieval results, dOTF can be compared to it as truth and qualified at the highest technology readiness levels.

*Address all correspondence to: Johanan L. Codona, E-mail: jlcodona@gmail.com

## 2 Differential OTF Wavefront Sensing

In Fourier optics,[14] an optical system is described by its pupil stop, the wavelength, and any aberrations and vignetting described as if they occurred in the pupil plane. If the light source is an unresolved point (i.e., a star), the resulting focal plane intensity is the point spread function (PSF). We write the pupil field as $\psi(\mathbf{x}) = \Pi(\mathbf{x})\psi_0(\mathbf{x})$, where $\mathbf{x}$ is the vector position in the pupil plane, $\Pi(\mathbf{x})$ is the (possibly complex) pupil mask, and $\psi_0(\mathbf{x}) = \alpha(\mathbf{x})\exp\{i\phi(\mathbf{x})\}$ is the field including any phase ($\phi$) and amplitude ($\alpha$) variations. The Fourier transform of the PSF is called the OTF[14,15] and can be written as the spatial average of the mutual coherence function over the pupil plane:

$$\mathcal{O}(\xi) = \int \psi(\mathbf{x}' + \xi/2)\psi^*(\mathbf{x}' - \xi/2)\mathrm{d}^2x'. \tag{1}$$

This is more concisely written in terms of correlations:

$$\mathcal{O} = \psi \star \psi^*, \tag{2}$$

where the correlation operator is defined as

$$f \star g = \int f(\mathbf{x}' + \xi/2)g(\mathbf{x}' - \xi/2)\mathrm{d}^2x'. \tag{3}$$

Introducing a localized modification in the pupil mask, $\Pi \to \Pi + \delta\Pi$, changes the pupil field $\psi = \Pi\psi_0 \to \psi + \delta\psi = \psi + \delta\Pi\psi_0$, causing the OTF to change by[10,16]

$$\delta\mathcal{O} = (\psi + \delta\psi)\star(\psi + \delta\psi)^* - \psi\star\psi^*, \tag{4}$$

$$= \psi\star\delta\psi^* + \delta\psi\star\psi^* + \delta\psi\star\delta\psi^*. \tag{5}$$

If $\delta\psi$ is sufficiently localized that it acts like a delta function, the dOTF becomes

$$\delta\mathcal{O}(\xi) \approx \mu^*\psi(x_0 + \xi) + \mu\psi^*(x_0 - \xi) + \delta\psi_+ \star \delta\psi_-^*, \tag{6}$$

where $\mu = \psi_0(\mathbf{x}_0)\int\delta\Pi(\mathbf{x}')\mathrm{d}^2x'$. This gives the pupil field which is offset by the location of the pupil modification, along with a conjugated and reflected version which is offset in the opposite direction. The quadratic term is localized around $\xi = 0$ and always lies within the overlap of the two complex pupil field images. So long as the pupil modification is introduced near the edge of the pupil, the overlap region between the two linear field terms will be small, allowing us to study the majority of the pupil field's phase and amplitude directly. Making a second modification elsewhere in the pupil will produce a similar field map, but with the new field modification's location obscured instead. If the change is insufficiently localized to use Eq. (6), the full equation, Eq. (5), gives a similar result, but with the complex pupil field cross-correlated with the complex conjugated difference field $\delta\psi^*$, somewhat blurring the result. Since pupil modifications usually have to involve a large area in order to get a high-quality signal and certainly are in our proposed JWST operations, we will use the full dOTF equation Eq. (5) while thinking of Eq. (6) as a conceptual idealization. The full dOTF is most clearly written as

$$\delta\mathcal{O}(\xi) = \delta\mathcal{O}_+(\xi) + \delta\mathcal{O}_+^*(-\xi) + \delta\mathcal{O}_{\delta\delta}(\xi), \tag{7}$$

where $\delta\mathcal{O}_+(\xi) = \psi\star\delta\psi^*$ and $\delta\mathcal{O}_{\delta\delta}(\xi) = \delta\psi\star\delta\psi^*$.

Making a dOTF measurement is extremely simple. Take two images of a star, one with the unmodified pupil and the other with some change introduced near the pupil's edge. Fourier transform the star images and subtract. The result is the dOTF, which gives us an estimate of the complex pupil field, albeit blurred by integration with the difference field introduced during the pupil modification. We also note that the star images must not be saturated and should be Nyquist-sampled over the entire filter band to avoid aliasing in the dOTF pupil.[11]

If the telescope shows a field position dependent PSF caused by non-pupil-plane aberrations or vignetting, we can still study it using the dOTF. In all cases, the PSF is a quadratic functional of the pupil field and the response to a pupil modification can be analyzed using a functional derivative of the more complete optical model.[16] We will not need to do that here, but in the practical case of a position-dependent PSF that does not change significantly over the size of the PSF, we can simply process two star images at various camera field positions in the same way described above. This will result in a set of position-dependent dOTFs, each one giving an estimate of the apparent pupil field that is backprojected to the pupil from the location of the star image on the sensor, including any non-common-path aberrations, vignetting, or other out-of-pupil-plane variations along the intervening ray paths.[12,13] Measurement of the dOTF can be made at multiple focal plane locations simultaneously, also with just two images, by looking at a field of stars.[12,13] The individual stars' halos will need to be sufficiently separated that they do not overlap. All stars in the image will be affected by the pupil modification, and the resulting collection of pupil fields projected along the various lines of sight can give stereoscopic or even tomographic information about the three-dimensional distribution of aberrations within the telescope. The dOTF theory[11] also describes the effect of optical bandwidth, which causes a proportional radial blurring of the pupil field about the point of modification at an amount equal to the fractional bandwidth (i.e., $\delta\xi/\xi = \delta\lambda/\lambda$). Narrower fractional bandwidths are preferred simply to avoid the radial blurring and are readily available with the JWST filter choices.

## 3 Using dOTF with JWST

The JWST design and construction are well past the point where any new changes to hardware or systems may be introduced. However, this does not necessarily preclude new applications for existing sensors and mechanisms. We propose two methods for using dOTF to measure the JWST primary mirror's OPD, including non-common-path aberrations and vignetting to the three imaging cameras. This technique can provide the information needed to fine-align and co-phase the primary segments, calibrate the non-common-path aberrations to the various cameras, as well as safely monitor image quality during normal operations. The first of these techniques (Sec. 3.1) involves blocking a small portion of the pupil with a purposefully misaligned filter wheel. Since this technique requires position control of a pupil plane filter wheel outside of the normal alignment position, it is possible only with NIRCam. The more general technique (Sec. 3.2) involves making a small displacement of a single outer segment actuator and works with all three imaging cameras: NIRCam, MIRI, and NIRISS. To avoid aliasing in dOTF, each camera must use filters with long-enough wavelengths to be at least Nyquist-sampled with the sensor's pixels (i.e., plate scale $< \lambda/2D$). The minimum wavelength for each camera and band is as follows: NIRCam short wavelength (SW) (plate scale 32 mas) 2.0 $\mu$m; NIRCam long wavelength (LW) (plate scale 65 mas) 4.1 $\mu$m; MIRI (plate scale 110 mas) 6.9 $\mu$m; and NIRISS (plate scale 65 mas) 4.1 $\mu$m. The shorter filter wavelengths give greater accuracy, while longer wavelengths allow for more coarse alignment with a larger capture range before phase wrapping occurs. By combining dOTF measurements in two longer bands, it is possible to uniquely determine phase jumps over multiple wavelengths, greatly increasing the capture range for coarse adjustment (Sec. 3.4). These considerations are common to both the pupil blocking and segment adjustment methods.

### 3.1 *Filter Wheel Method*

The NIRcam filter wheels[17,18] are driven in such a way that they can be arbitrarily positioned outside of their nominal placements. As the filter wheel is rotated, the circular filter mount will first obscure an edge of one of the outer segments. So long as the step size and placement accuracy are at most a few percent of the pupil diameter, the resulting obscuration should make an ideal pupil modification. Since the filter used to perform this blocking must be in the pupil plane as well as be Nyquist-sampled, it will only be possible with the long wavelength, pupil plane (LWP) filters F405N, F418N, F466N, and F470N (Table 1). Since MIRI includes a ratchet mechanism in its filter wheel assembly and the only Nyquist-sampled filters in NIRISS are in the focal plane, the filter wheel method cannot be used with either of these cameras.

**Table 1** Usable filters for differential optical transfer function (dOTF) measurements. This table summarizes the filters that best satisfy the requirement that the point spread function be Nyquist-sampled and minimally blurred radially by the filter bandwidth. Nyquist sampling is when there are 2 or more pixels across a $\lambda/D$ spot in the field. The usable near-infrared imager and slitless spectrograph (NIRISS) filters require The clear pupil wheel mask (CLEARP) filter to be selected, which somewhat obscures parts of the pupil. Three MIRI filters are four-quadrant coronagraphic phase masks. These should still work so long as the star is placed sufficiently far from the phase shift boundaries. The best accuracy will be achieved using NIRCam's narrowband filters F212N or F225N, or the 10% bandwidth F210M. The best single-band capture range before phase wrapping will be using MIRI's coronagraphic phase masks. By making dOTF measurements in two bands and comparing the difference phase, a much larger range is possible before wrapping. The greatest two-band range (58.5 $\mu$m) is achieved by using NIRCam's F466N combined with F470N. The accuracy listed is scaled by wavelength from our experiment, but is actually more complicated. See Sec. 6 for more detail.

| Camera | Filter | Location | $\lambda$ ($\mu$m) | $\delta\lambda/\lambda$ (%) | Plate scale (mas/pixel) | Pixels/($\lambda/D$) | Accuracy (nm) | Phase wrap range ($\mu$m) | Comments |
|---|---|---|---|---|---|---|---|---|---|
| NIRCam | F210M | Short wavelength, focal plane (SWF) | 2.10 | 10 | 32 | 2.1 | 32 | 1.05 | Best accuracy |
| | F430M | Long wavelength, focal plane (LWF) | 4.30 | 4.7 | 65 | 2.1 | 65 | 2.15 | |
| | F460M | LWF | 4.60 | 4.3 | 65 | 2.2 | 69 | 2.30 | |
| | F480M | LWF | 4.80 | 8.3 | 65 | 2.2 | 72 | 2.40 | |
| | F212N | SWF | 2.12 | 1 | 32 | 2.1 | 32 | 1.06 | Narrow-band (NB) filter; best accuracy |
| | F225N | Short wavelength, pupil plane | 2.25 | 1 | 32 | 2.2 | 34 | 1.12 | NB filter; filter wheel method |
| | F405N | LWP | 4.05 | 1 | 65 | 2.0 | 61 | 2.03 | NB filter; filter wheel method |
| | F418N | LWP | 4.18 | 1 | 65 | 2.0 | 63 | 2.09 | NB filter; filter wheel method |
| | F466N | LWP | 4.66 | 1 | 65 | 2.3 | 70 | 2.33 | NB filter; filter wheel method |
| | F470N | LWP | 4.70 | 1 | 65 | 2.3 | 71 | 2.35 | NB filter; filter wheel method |
| MIRI | F1130W | P | 11.3 | 6 | 110 | 3.3 | 170 | 5.7 | |
| | F1065C | MIRIM | 10.65 | 5 | 110 | 3.1 | 160 | 5.3 | 4QPM_1 |
| | F1140C | MIRIM | 11.4 | 5 | 110 | 3.3 | 171 | 5.7 | 4QPM_2 |
| | F1550C | MIRIM | 15.5 | 5 | 110 | 4.5 | 233 | 7.8 | 4QPM_3 |
| NIRISS | F430M | F | 4.3 | 5 | 65 | 2.1 | 65 | 2.2 | Requires CLEARP |
| | F480M | F | 4.8 | 6 | 65 | 2.3 | 72 | 2.4 | Requires CLEARP |

Figure 1 shows a simulation of the NIRCam filter wheel blocking method using the F212N narrowband filter. Figure 1(a) shows the OPD map used in the simulation, selected from the 155 nm rms WFE set in JWST PSF simulator WebbPSF.[19] Since filter F212N is in the NIRCam SW band, the plate scale is 32 mas, resulting in the star image shown in Fig. 1(b). Fourier transforming the star image gives a complex OTF [Fig. 1(c)], which does not readily show any useful phase information in this case. After blocking the edge of the leftmost segment and collecting a second image and OTF (not shown), we compute the dOTF in Fig. 1(d). The pupil modification always appears in the middle of the dOTF, indicating that the pupil image on the right hand side of the dOTF corresponds with the OPD mask orientation in Fig. 1(a). Since the phase excursion is small, two-dimensional phase unwrapping is unnecessary and can be directly used to estimate the primary surface height using $h = \arg\{\delta\mathcal{O}\}\lambda/4\pi$.

Each side of the dOTF gives the complex pupil field cross-correlated with the conjugate of the complex field change over the obscured segment area. For measuring segment pistons, the correlation causes a blurring in the direction of the obscured segment's edge, but is minimal in the segment center. If we can accept the relatively small amount of blurring in the rotation direction of the filter wheel, the transverse blurring will appear essentially one-dimensional and deconvolution will be simplified.

## 3.2 *Moving an Outer Segment*

It is possible to introduce a pupil modification in a single segment by moving one of the outer support actuators by a small distance. This method will work with all three imaging cameras and allows not only segment co-phasing, but also separate determination of the non-common-path aberrations between the various cameras. Due to phase wrapping effects in $\delta\psi$, the maximum recommended displacement of the moved actuator is $\lambda/4$, but smaller displacements will work as well.[11] For the shortest wavelength usable for dOTF with NIRcam, 2.1 $\mu$m (Table 1), a maximum 500 nm displacement is recommended, but a 100 nm (or smaller) displacement can also be effective with appropriate exposure and target star magnitude choices. In contrast to moving an entire segment in piston, moving a segment in both tip-tilt and piston such that an edge adjacent to the other segments remains stationary will produce a field change over an area that is smaller than the entire segment. This will minimize, but not eliminate, the blurring. Any segment in the outer ring can be used to make the measurement, but the six outermost corner segments will introduce the least pupil overlap.

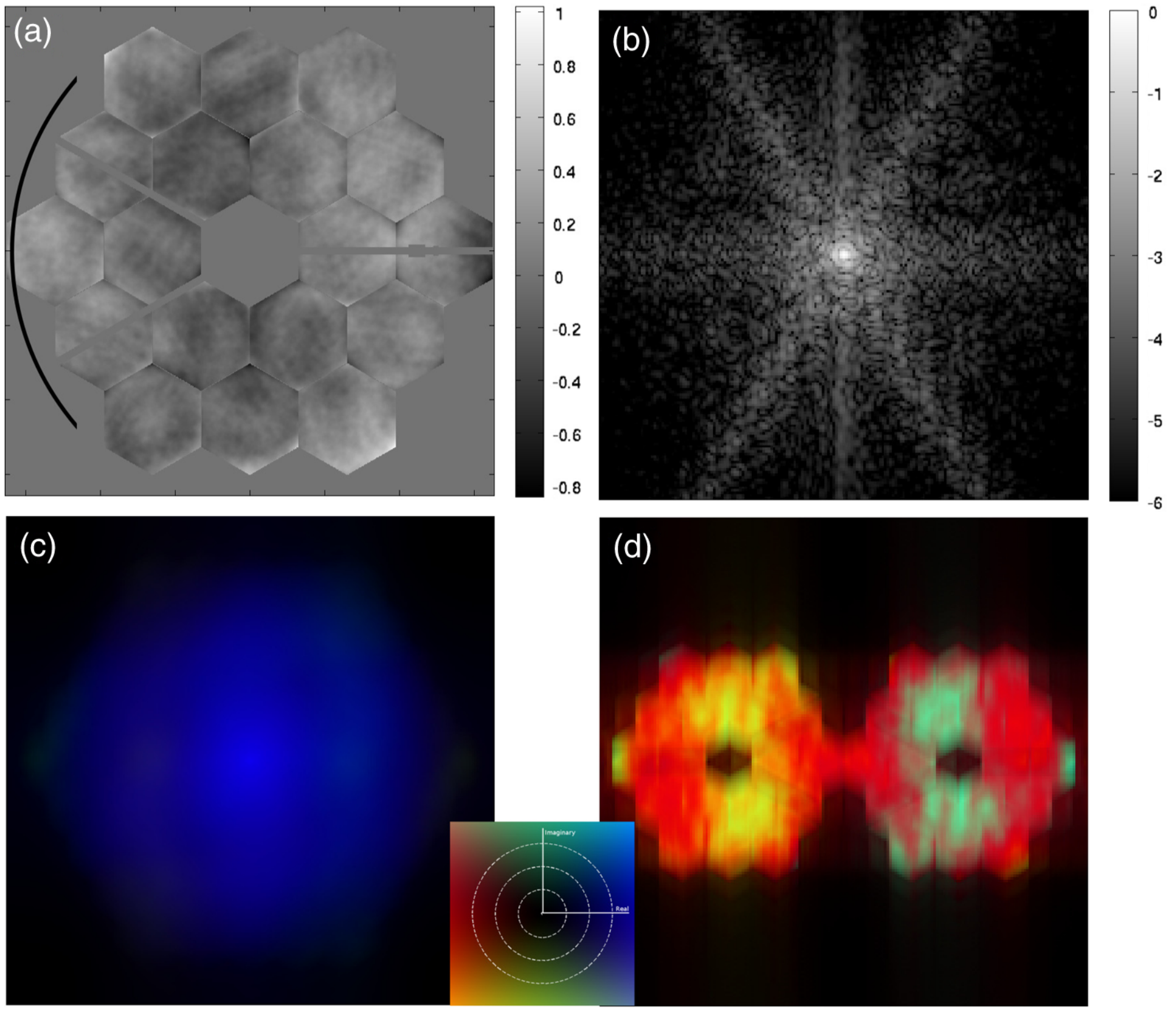

**Fig. 1** Simulation of the filter wheel blocking method. (a) shows the OPD map with a 161.5 nm rms wavefront error. The black arc is a schematic for the edge of the filter holder, adjusted so that the leftmost segment is slightly obscured. The NIRCam F212N image of a star is shown in (b) and the corresponding complex optical transfer function (OTF) in (c). After blocking the edge of the leftmost segment and collecting a second star image and OTF (not shown), we compute the differential OTF (dOTF) in (d). The complex functions in this and other figures in this paper are plotted using the inset color map: brightness corresponds to amplitude, while color represents phase (blue/red are $+/-$ real, while green/yellow are $+/-$ imaginary).

A simulation of this method is shown in Fig. 2 with the same choices made as in Fig. 1. The same test OPD map was used [Fig. 1(a)], but a small tilt was applied to the segment marked by * such that the maximum displacement was 75 nm [Fig. 2(b)]. Such a small displacement would only alter the overall rms WFE by at most <20 nm, and possibly much less. Therefore, the telescope will not be considered to be performance impaired even in the modified configuration. Since a small OPD change affects the PSF less than a larger one, the dOTF signal will be smaller, requiring a longer exposure to compensate (Sec. 6). As before, the location of the pupil modification appears in the center of the dOTF and the two sides are conjugates of each other providing redundant information. Since the overall phase is not measurable through the PSF intensity, the overall phase of the dOTF will vary and will be different depending on the nature, location, and details of the pupil modification. To more easily compare the two simulated methods, we applied an overall phase shift to the computed dOTF to better compare with Fig. 1(d). Since this operation does not preserve the conjugate symmetry to both sides of the dOTF, only the intended side is to be compared. Since the segment tilt method involves a larger area than the filter wheel obscuration assumed in Sec. 3.1, the result in Fig. 1(d) shows less fine detail. However, both techniques are good enough to significantly improve the OPD of the mirror without further processing.

### 3.3 *Difference Field and Bandwidth Effects*

Larger actuator displacements will result in more phase wrapping in the difference field, causing it to have finer structure, which may be useful for studying smaller pupil field structures.[11] Even with more modest actuator displacements, the cross-correlation will couple pupil field phase variations into the dOTF amplitude, which itself can be a powerful tool for dialing-out phase discontinuities. That is to say, even though the amplitude of the pupil field is essentially constant across a wavefront discontinuity, the cross-correlation with the conjugate of the difference field can cause an obvious change in the dOTF amplitude, which will vanish when the segments are properly aligned. Designing delta fields to fully exploit this effect for telescope wavefront control will require further study. Another area for further study is the potential for deconvolution of the dOTF into the original pupil field and the difference field. Even though this may not be necessary for basic segment phasing since the dOTF shows phase and phase discontinuities clearly enough to adjust them, deconvolution may still be desired. Since we will have a good idea of how much we moved a test segment as well as the fact that the deconvolved field should have a nearly uniform amplitude on the segments and be constrained by the pupil aperture, there is a positive outlook for extracting an unblurred pupil field map from a single dOTF measurement. This is a subject of continuing study.

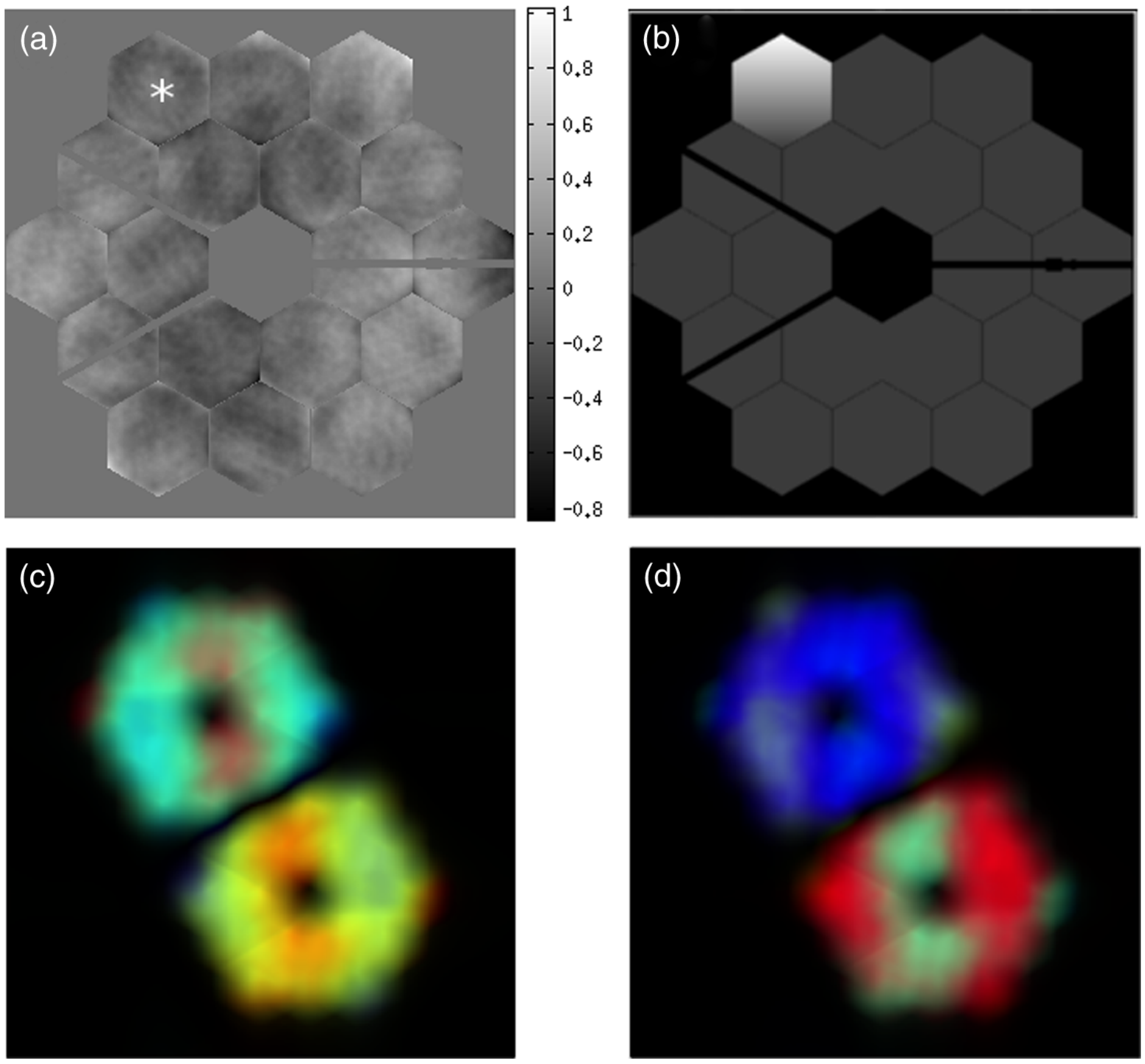

**Fig. 2** Simulation of dOTF by segment tilting. Image (a) shows the OPD map—the same as in Fig. 1(a). The segment used to make the pupil modification is marked with an asterisk "*". The segment tilt is shown in context in (b). A baseline star image was computed along with a second one including the tilted segment. The difference of the Fourier transforms gives the dOTF shown in (c). The pupil field corresponding to (a) is in the lower-right part of (c). Although it alters the conjugated side, we can multiply the dOTF in (c) by an overall phase shift to better compare with the right hand side of the dOTF in Fig. 1(d).

The JWST cameras provide multiple filter choices for narrowband dOTF measurements (Table 1), so using bandwidths broader than 10% should not be required. However, it is useful to see what happens if we were to use a broadband filter to make a dOTF measurement. Figure 3 shows two random simulations, one measured using a 10% filter bandwidth and the other with a 50% bandwidth. The complex pupil fields [Figs. 3(a) and 3(d)] are shown as complex so they can be directly compared with the dOTF results [Figs. 3(c) and 3(f)]. The familiar radial chromatic blurring of the diffraction pattern in Figs. 3(b) and 3(e) appears in reverse in the dOTFs, with the dOTF radially blurred away from the location of the pupil modification. The amount of blurring is proportional to the distance by an amount $\delta r/r = \delta\lambda/\lambda$. The 10% case is only marginally worse than the narrowband simulations in Fig. 2 and could just as easily be used without further processing. The 50% case would perhaps present a more significant loss of quantitative accuracy, but it is still easily interpreted and clearly shows the misaligned segments and their offset magnitudes. The conclusion is that the filter bandwidth is not going to be a significant issue for JWST dOTF measurements.

### 3.4 *Phase Wrapping Range and Combining Bands*

Since the dOTF wraps in phase, similar to an interferometric measurement of the pupil field, it cannot measure wavefront phase jumps larger than $2\pi$ (i.e., mirror surface jumps larger than $\lambda/2$). Since the wavefront's phase includes a down-and-back travel distance to the primary mirror, the measurement loses the ability to determine segment piston differences greater than $\lambda/2$ unless there is a ramp connecting various levels in the OPD map. Large intersegment jumps will be subject to this problem. For the various usable NIRCam filters, the phase wrapping occurs for jumps >1.05 to 2.35 $\mu$m of piston. The MIRI phase wrapping limits the range from 5.3 to 7.8 $\mu$m and NIRISS is either 2.2 or 2.4 $\mu$m (Table 1). These limits are more than adequate for the anticipated fine alignment needs of the JWST, but could be of more use during coarse alignment if the capture range were larger.

There is a simple method for extending the phase wrapping range that uses two sets of dOTF measurements in two different wavelength bands. The idea is similar to using two wavelength interferometry,[20] which extends the phase wrapping from an OPD of the single wavelength $\lambda$ to a greatly extended range of $\Lambda_{\mathrm{eff}} = \lambda_1\lambda_2/|\lambda_2 - \lambda_1|$ with two measured wavelengths. By measuring the dOTF at two close wavelengths, say using NIRCam with filters F466N and F470N, the phase wrapping limit increases from a mirror surface displacement of 2.35 to 110.6 $\mu$m. This pairing results in such a large increase in range because the filters are quite close together in wavelength. Combining the MIRI filters F1130W with F1140C (one of the MIRI imager (MIRIM) four-quadrant phase masks, but still usable with dOTF), it becomes possible to uniquely measure segment displacements up to 0.638 mm, albeit with a lower accuracy. The dOTF phase difference can be measured without introducing phase wrapping artifacts on the individual filters' wrapping scales by computing $\delta\phi = \arg\{\delta\mathcal{O}_1(\xi)\delta\mathcal{O}_2^*(\xi)\}$ (Fig. 4). Using this difference technique combined with the individual band dOTFs allows the OPD map to be measured, including large intersegment jumps, with a large dynamic range at the accuracy of a single wavelength measurement.

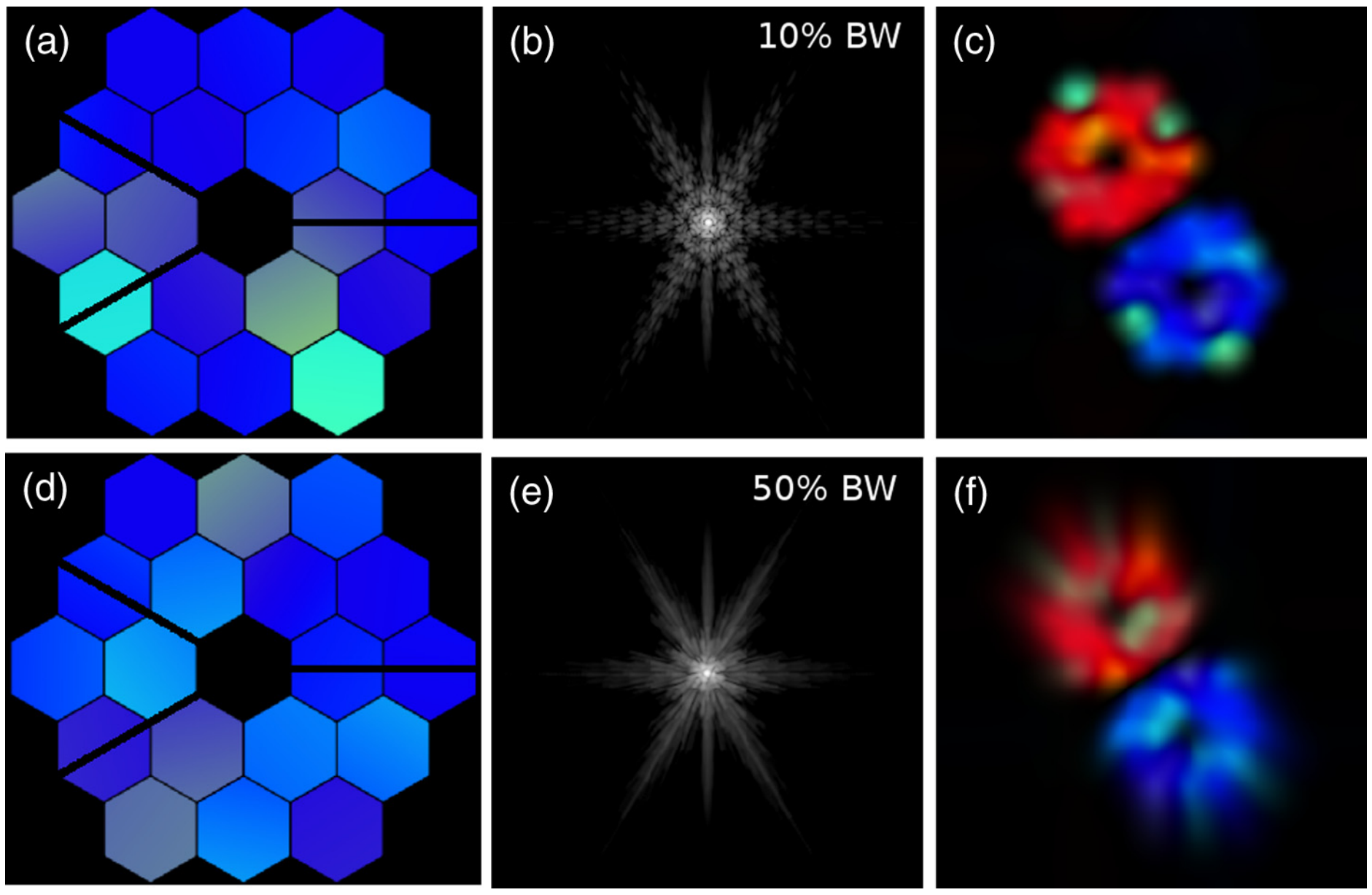

**Fig. 3** The effect of larger bandwidths on segment-tilt dOTF measurements. Two random simulations are shown, one with a 10% bandwidth [(a) to (c)] and the other with a 50% bandwidth [(d) to (f)]. For better display, an overall (unobservable) phase has been included to rotate the lower-right halves of the complex dOTF to roughly match the phases of the inputs. The radial bandwidth smearing in the point spread functions (PSFs) (b) and (e) lead to radial smearing of the complex dOTF with an effect that is proportional to the distance from the actuated segment. Since there are many narrowband filter options available, extreme broadband measurements should not be necessary. However, this simulation clearly shows that even without any corrections, a broadband dOTF gives a useful pupil field estimate.

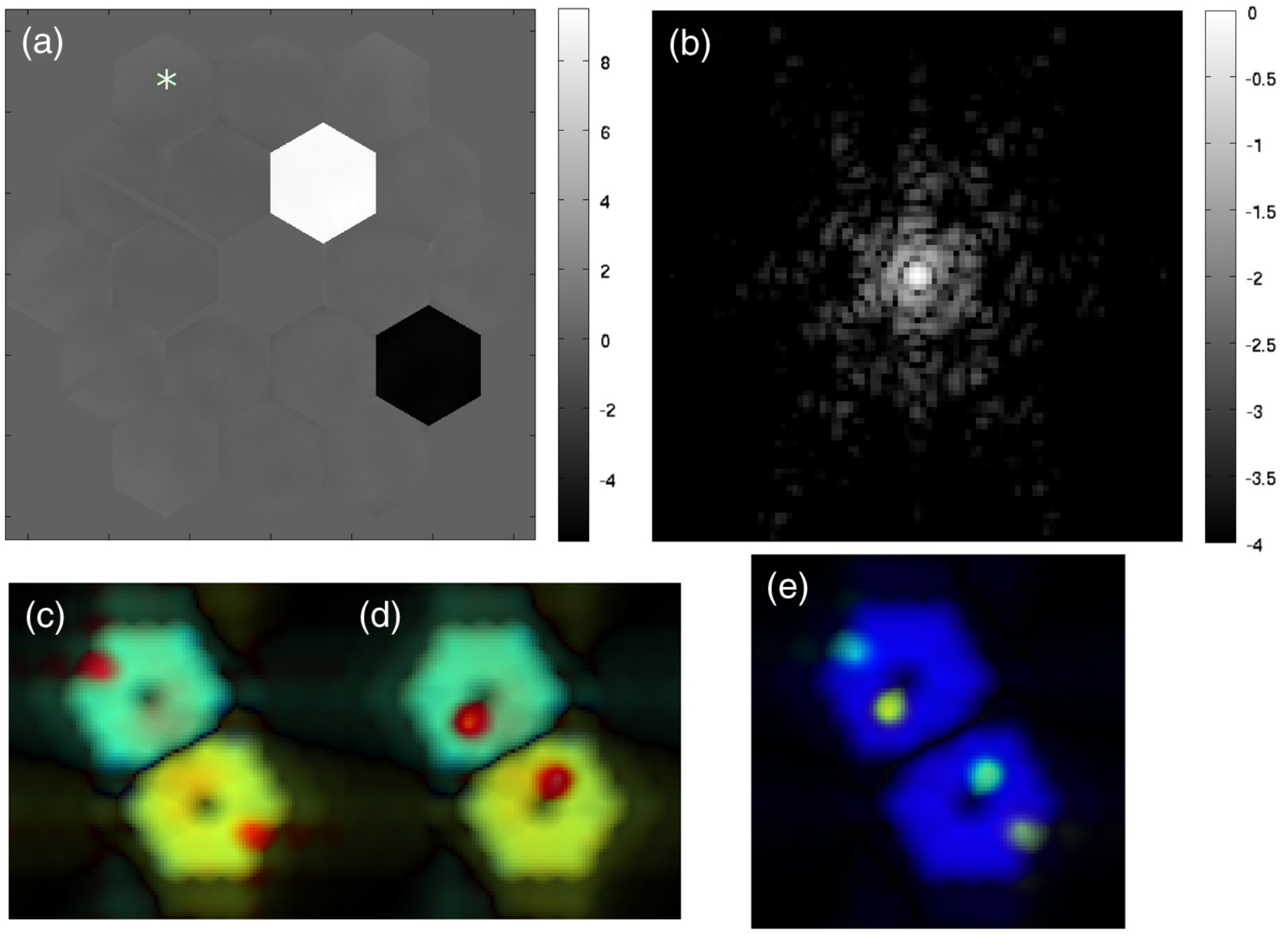

**Fig. 4** An example of using dOTF in two wavelength bands to measure large OPD jumps. The test OPD is shown in (a) as one of the WebbPSF sample OPD maps combined with a very large piston error in two of the segments. The upper (white) segment is displaced upward by 4.18 $\mu$m, while the lower (black) segment is displaced downward by 2.35 $\mu$m. These values were chosen to cause a multiple of a full wavelength of phase shift in each of our chosen NIRCam filter bands: F418N and F470N. The upper-left segment (marked by *) was tilted by 75 nm in order to make each dOTF measurement, four images used in total. An example narrowband star image is shown in (b), with 3 arcsec on a side and a plate scale of 65 mas. Since the PSF size scales with wavelength, the dOTF scales inversely. This is easily corrected during the Fourier transforms, allowing us to compute the dOTFs to the same spatial scale. The dOTF for filter F418N is shown in (c), while that for F470N is shown in (d). Notice that each shows only one of the displaced segments, while the other wraps in phase to a multiple of $2\pi$. The visible phase structures in (c) and (d) are appropriate for the individual bands, but are not able to measure the much larger displacements. In (e), we see the result of plotting $\delta\mathcal{O}_{\mathrm{F418N}}\delta\mathcal{O}^*_{\mathrm{F470N}}$, which wraps more slowly after a displacement of 16.8 $\mu$m. The heights of the two displaced segments are clearly visible with quantitative displacement values.

## 4 Validating Experiments

To verify that this new technique will indeed work as expected with a segmented mirror, we performed a series of experiments. We measured the dOTF using a small pupil obstruction as well as by changing a single outlying segment of a hexagonal deformable mirror (DM). We performed our experiments with the New England College of Optometry adaptive optics fundus camera[21] (Fig. 5), later moved to the College of Optometry at Ohio State University. For these tests, the ophthalmic imaging system can be thought of as a telescope imaging an unresolved point source placed in what would normally be the subject's retinal plane. The model eye used here is a pupil lens adjusted to collimate a 633 nm fiber-coupled laser point source in the retina plane. From here, the pupil lens is conjugated to a 37-segment Iris AO MEMS DM.[22] The light was then reimaged onto a 3.5 mm pupil iris that defines the system aperture and was adjusted to partially obscure the outer ring of DM segments during the experiment. The light was then imaged onto a Rolera MGi PLUS camera where the PSFs were recorded to 14 bits.

For the first test reported here, the DM was removed and replaced with a plane mirror. The pupil modification was a 25-gauge [0.5144 mm diameter (i.e., $0.15D$)] hypodermic needle inserted into the beam just inside the pupil edge and very near to the pupil plane. We used 30 co-added image frames with the needle completely outside the pupil as a baseline PSF. The needle was then moved inside the edge of the pupil at various distances with 30 co-added images taken at each position (Fig. 6). Three of the resulting dOTFs are shown in Fig. 7. The wavefront aberration over the pupil is easily seen and measured, indicating a Strehl ratio of 92%. As the needle was inserted further into the pupil, the dOTF signal increased, as did the blurring due to the autocorrelation, while the OPD shape stayed essentially unchanged. This behavior was exactly as expected from the dOTF theory.

We also used the segmented DM (Fig. 8) to both create a test pattern and introduce a phase-only pupil modification. The DM was flattened to the factory reference and an increasing piston of 25 nm steps was added to the first seven segments (Fig. 8). The DM's outer ring of segments was partially obscured by the 3.5 mm circular pupil mask, so the convolving field difference was smaller than a normal full segment. In Fig. 9, we show one of the dOTF results based on a 1 mrad tilt applied to segment 22 (see Fig. 8). The measured phases were consistent with the applied test pattern, with a DM surface height resolution in our test of better than 10 nm averaged over individual segments. Although it will probably be possible to take better dOTF measurements than those with the JWST, we use this experiment to estimate the potential measurement accuracy with the JWST. Since 10 nm is $\sim$1.5% of our test wavelength, we will use $0.015\lambda$ as a rough estimate of dOTF measurement accuracy in Table 1. In reality, the accuracy will depend on wavelength, plate scale, camera characteristics, background noise, etc. We examine this issue for an example case in Sec. 6.

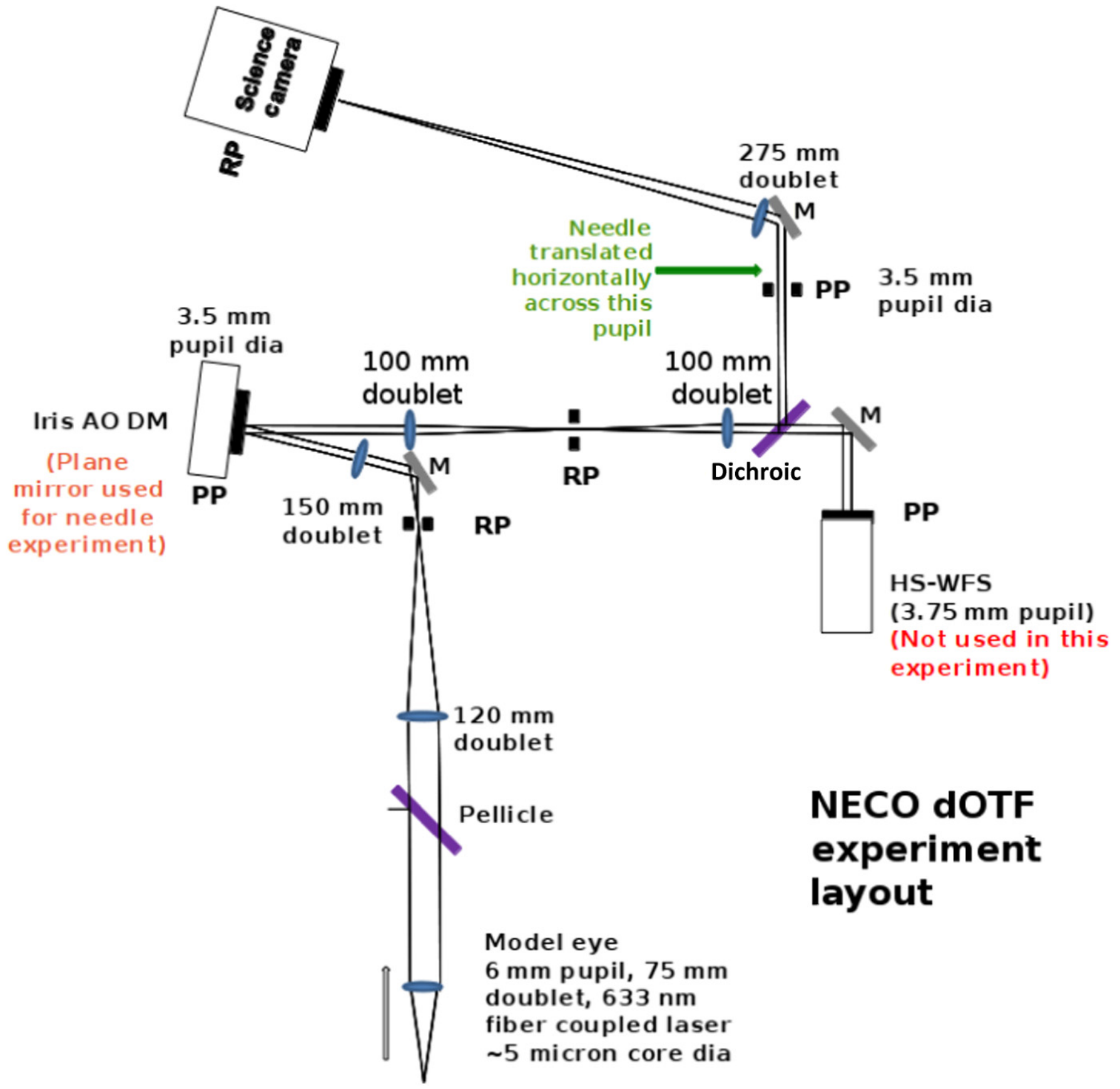

**Fig. 5** New England College of Optometry (NECO) experimental setup for the segmented mirror tests.

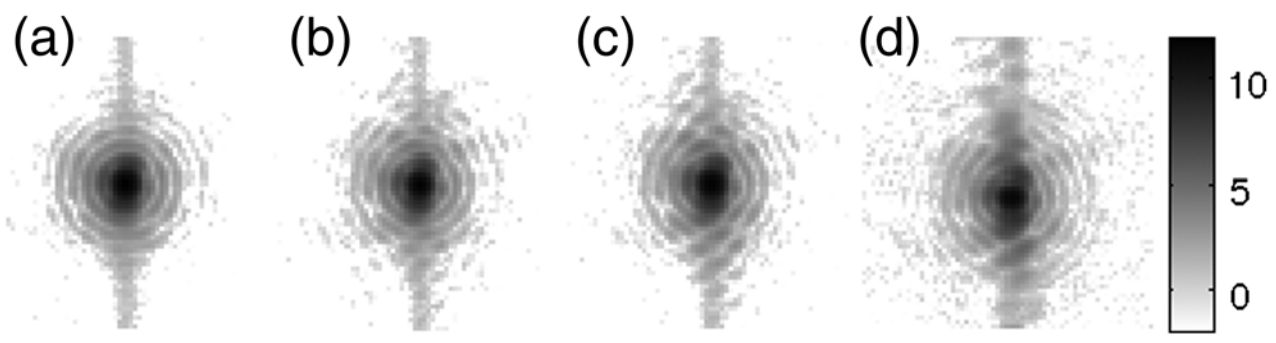

**Fig. 6** NECO needle test PSFs. The negative logarithmic gray-scale image is labeled in bits. The PSF peaks are unsaturated and adjusted to be approximately half a bit below the full well. The vertical bleeding of the PSF core is at a very low level and appears in the corresponding OTFs as a single short row of pixels with enhanced noise. Because of the location of the probe needle relative to the readout direction of the sensor, the noise falls in the dOTF overlap region, or beyond the pupil limits. Therefore, it has no impact on the observations. The full-pupil PSF is shown in (a) while the remaining three PSFs resulted when inserting a needle into the pupil edge by (b) 0.1 mm, (c) 0.5 mm, and (d) 2.0 mm. The corresponding dOTF measurements are shown in Fig. 7.

## 5 Usable JWST Filters for dOTF

The various filters and cameras usable for JWST dOTF measurements are summarized in Table 1. The filters were selected for Nyquist sampling with the corresponding sensor and configuration, and limited to filter bandwidths <10%. Nyquist sampling requires at least 2 pixels per $\lambda/D$. The measurement accuracy is based on the lab experiments that achieved wavefront accuracies of 1.5% $\lambda$ and does not take into account important details such as photon noise and camera characteristics. We see in Sec. 6 that the actual accuracy can vary by more than a decade due to noise statistics and exposure time alone. The accuracy value is meant as a somewhat conservative guide and may well be improved with experience and enhanced processing. The phase wrapping limit for single-wavelength measurements is shown as height differences of the mirror segments (i.e., one half the OPD or wavefront jump limit). As described in Sec. 3.4, measurements in two bands can be combined to measure much larger jumps. Of the four acceptable MIRI filters, three are 4-quadrant phase masks in the MIRIM coronagraph. This should not present a problem for use with dOTF so long as the star is not placed near the phase boundaries, but is at least $10\lambda/D$ in the clear. A Lyot stop should not be used with MIRIM dOTF measurements since it would serve no purpose and only obscure parts of the pupil that are being measured. The NIRISS filters require that the CLEARP mask be selected as the pupil filter. This obscures some of the pupil and contains an alignment

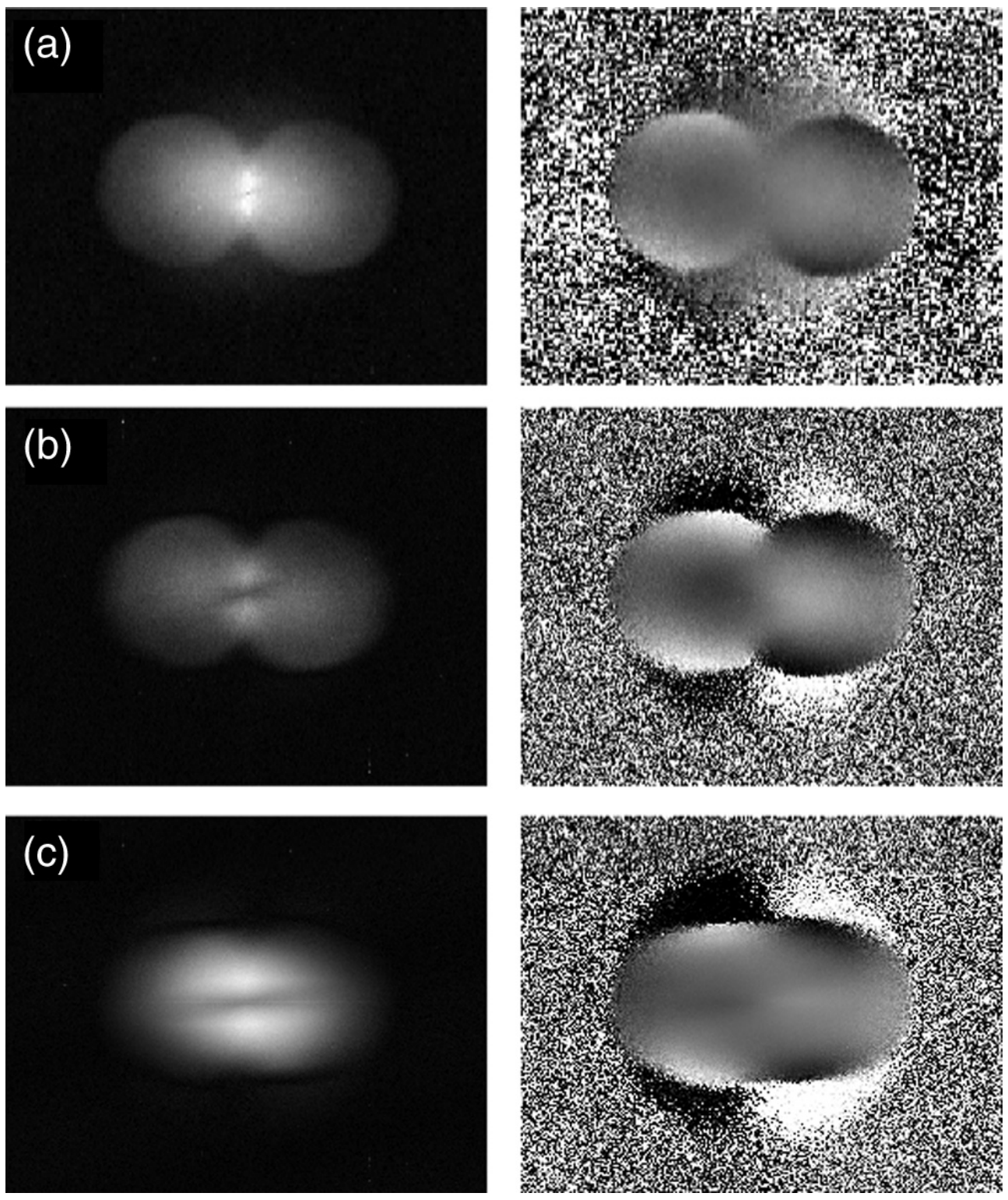

**Fig. 7** Pupil obstruction dOTFs. Each pair of images is the magnitude of the dOTF (left), $|\delta\mathcal{O}|$, and the phase of the dOTF (right), $\arg\{\delta\mathcal{O}\}$. Each dOTF was referenced to the same full pupil PSF image. In (a), the needle was placed ~0.1 mm inside the edge of the 3.5 mm pupil, resulting in a faint but clear dOTF signal. The image shown here was binned $2 \times 2$ in the complex dOTF to increase the S/N. In (b), the needle was inserted 0.5 mm, with a corresponding increase in signal as well as increased blurring. Finally, in (c), the needle is inserted 2 mm into the pupil (>50% of the the iris diameter), causing extreme blurring, as expected. The dOTF phase gray scale is adjusted to indicate a wavefront aberration of 300 nm from white to black.

mask, but should otherwise have no adverse effect on the dOTF measurements.

## 6 Example Noise Analysis

Photon noise causes actual measured dOTFs to be noisy estimates of the ideal dOTFs computed above. So long as nothing changes during an exposure, the estimated dOTF is the ideal value plus complex noise. Read noise and other noise sources appearing across the sensor lead to a complex Gaussian noise background, extending beyond the pupils into the regions with no dOTF power. The star image photon noise also contributes complex noise over the dOTF. The estimation error statistics can be determined from an ensemble of dOTF estimates derived from pairs of finite-exposure images. The image background noise causes a complex Gaussian noise across the entire dOTF image with little or no correlation between pixels, while the photon noise in the individual star images leads to complex noise which is correlated over regions comparable with the size of the pupil. A full noise analysis for dOTF measurements is beyond the scope of this paper, but we can benefit from a numerical study of the noise for a particular case.

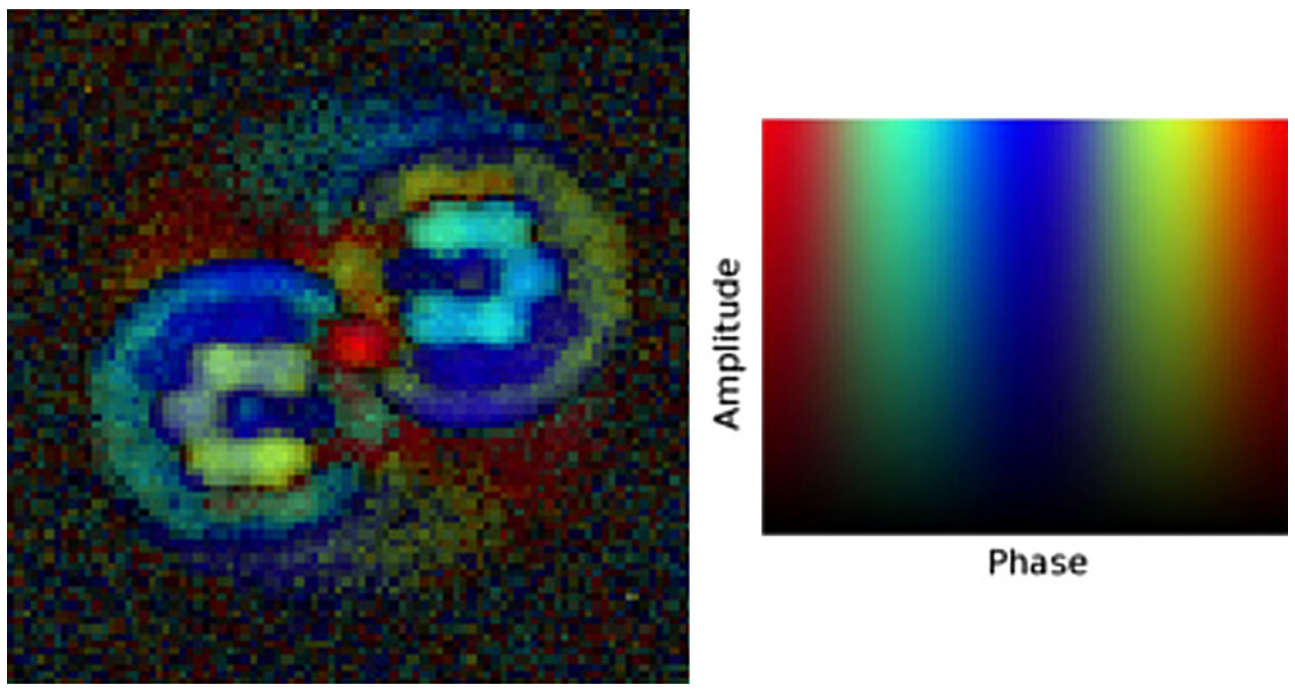

**Fig. 9** Plot of the measured complex dOTF. The DM was flattened and the first seven segments had 25 nm increasing piston steps applied as a test pattern. Segment 22, partially obscured by the overlaid circular pupil mask, was tilted by 1 mrad to create a phase-only pupil modification. No attempt was made to adjust the piston to make the modified OPD be continuous at the inner edge of the test segment as in the simulation in Sec. 3.2.

We will consider using NIRCam with the F466N narrowband filter. The plate scale is 65 mas or 2.3 pixels per $\lambda/D$. Assuming a well depth of 78,000 $e^-$, we adjust the exposure such that the brightest pixel is 92% of saturation, or $72{,}000\,e^-$ with a read noise of $10\,e^-$. This corresponds to an integrated count for the star of 545,000 $e^-$. Since the dark current is very low and we assume the star is imaged against a dark background, the dOTF noise depends only on how many photons are imaged, not the individual exposure times. As an example, assuming a quantum efficiency of 80%, an 8th magnitude star will require an exposure of 1.85 s. If we require more photons, we must take more exposures and co-add them for both the baseline and modified pupil. An example calculation is shown in Fig. 10. The sample OPD map is shown in Fig. 10(a), resulting in the complex field shown in Fig. 10(b) for $\lambda = 4.66$ $\mu$m. A $\log_2$ plot of a sample star image is shown in Fig. 10(c), including photon and read noise. Note that it will be possible to capture more photons in a single image at longer wavelengths without saturating since the PSF is spread over more pixels. The dOTF in Fig. 10(d) is computed after capturing a second exposure where a single segment [indicated by the * in Fig. 10(a)] is lifted by $\lambda/4$ or 1.15 $\mu$m. We used $95 \times 95$-pixel star images (6.175 arcsec) giving dOTF pixels of size 15.4 cm or an area equal to 1/60 of a single segment. We computed 1024 dOTF realizations for a selection of segment tilts and computed the complex standard deviation as

$$\sigma = \langle (\delta\mathcal{O} - \langle \delta\mathcal{O} \rangle)^2 \rangle^{1/2}. \tag{8}$$

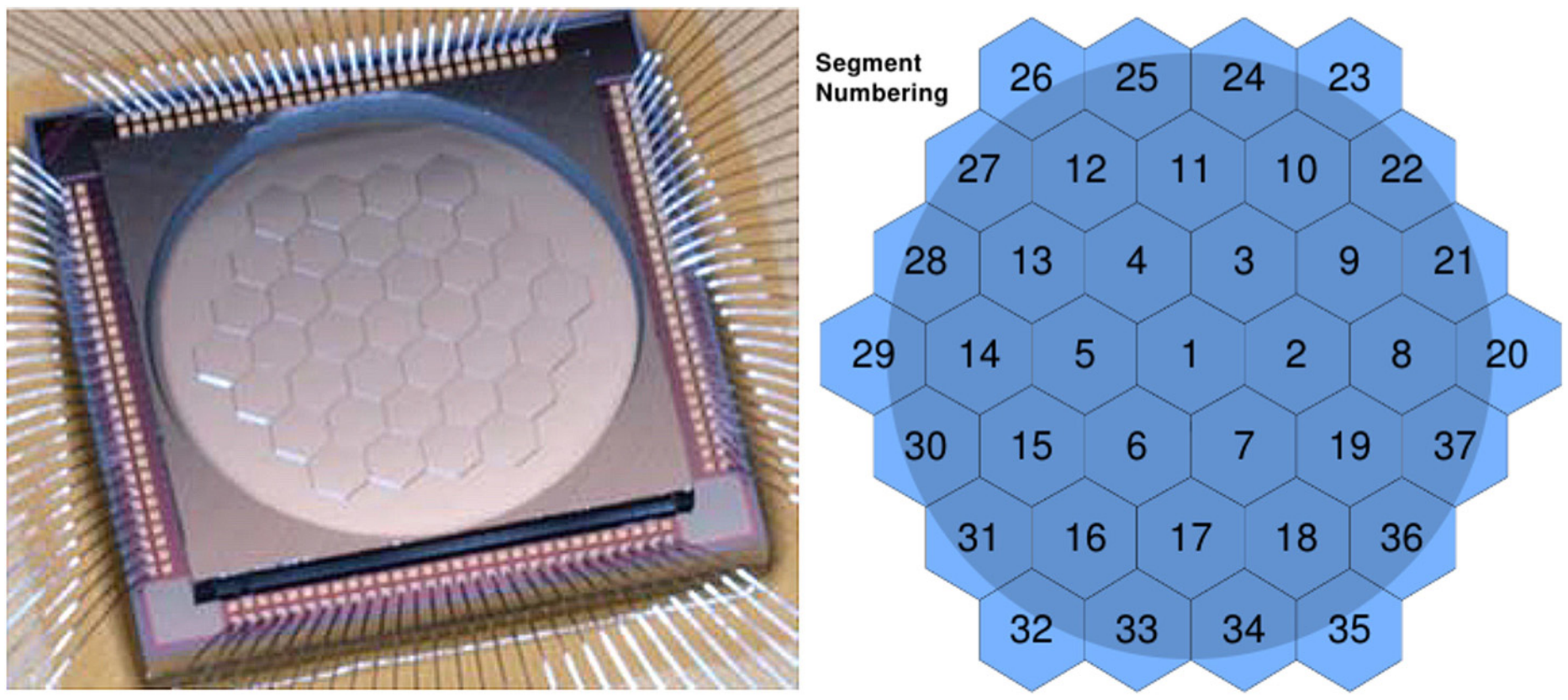

**Fig. 8** The Iris AO, Inc. PTT111 37-segment MEMS deformable mirror (DM)[22] used in the phase modification experiment. This DM itself has a hexagonal active area, but our experiment used an inscribed pupil mask that partially covered the outer ring of segments (shown as a darker circle on the right hand diagram).

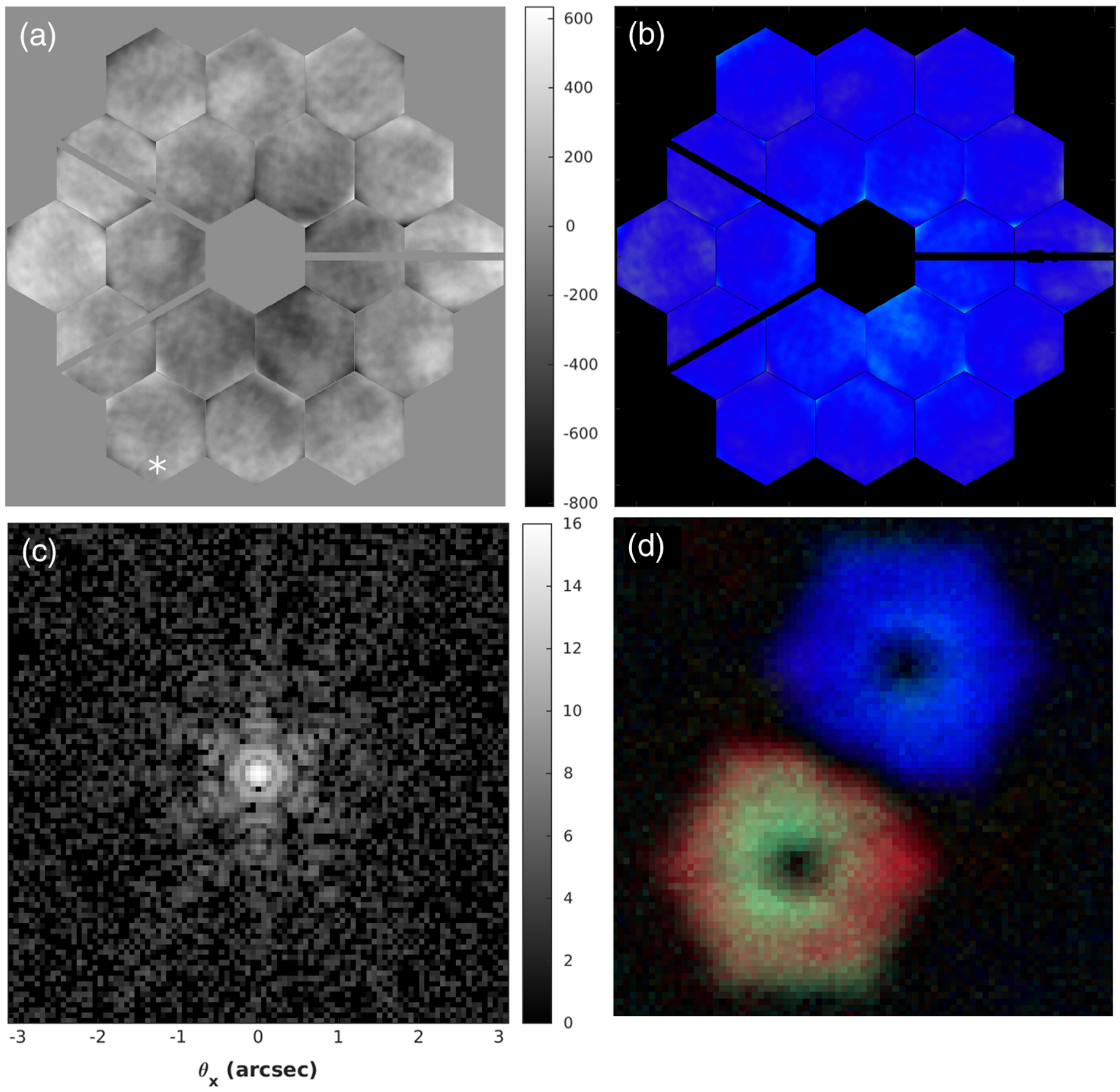

**Fig. 10** Example noise analysis for the tilted segment method using a mag 8 star with NIRCam F466N. The test OPD map is shown in (a), while the resulting complex amplitude at $\lambda = 4.66$ $\mu$m is shown in (b). An exposure time of 1.85 s brings the brightest pixel to 72,000 $e^-$ (92% of the 78,000 $e^-$ well depth), with a read noise of 10 $e^-$ resulting in image (c). Introducing a pupil modification by lifting the bottom edge of the lower-leftmost segment [marked * in (a)] by $\lambda/4$ (1150 nm), results in the dOTF estimate (d). The displayed dOTF was multiplied by a complex phase shift to make the upper-right portion more directly comparable with the pupil field in (b). The star image in (c) is 95 pixels on a side and the dOTF pixels in (d) are 15.4 cm on a side.

The central limit theorem suggests that the complex value of a given dOTF pixel is the mean value plus a zero-mean complex Gaussian error with a standard deviation of $\sigma$. We will interpret this as giving a phase error of

$$\epsilon_\phi = \arcsin(\sigma / |\langle \delta\mathcal{O} \rangle|) \tag{9}$$

or an OPD error of

$$\delta\text{OPD} = \lambda \epsilon_\phi / 2\pi. \tag{10}$$

The error in the surface height of the primary is $\delta\text{OPD}/2$. As a useful reference scale, we will consider the piston error measured over the area of an entire segment. If the noise in the dOTF pixels were statistically independent, we would expect that the error would be $\sqrt{60} = 7.7$ times smaller than the per-pixel error. However, since the photon noise from the star is correlated over larger spatial scales, the average piston error over a segment is more like 1.85 times smaller than the per-pixel value. This statistical piston error is shown in Fig. 11 for a variety of probe segment actuator motions and a number of co-added images. The maximum displacement of the lower edge of the * segment shown in Fig. 10(a) is 50, 125, 250, 500, and 1150 nm (the maximum recommended simple value of $\lambda/4$). Since a small segment displacement has only a small effect on the PSF, the dOTF signal is reduced, resulting in larger piston errors. If we collect more co-added images, we can compensate for a smaller segment motion. For an 8th magnitude star, acquiring two sets of 100 co-added images would take >6 min and give about the same accuracy measurements with a 75 nm displacement as a 4 s measurement using a $\lambda/4$ measurement.

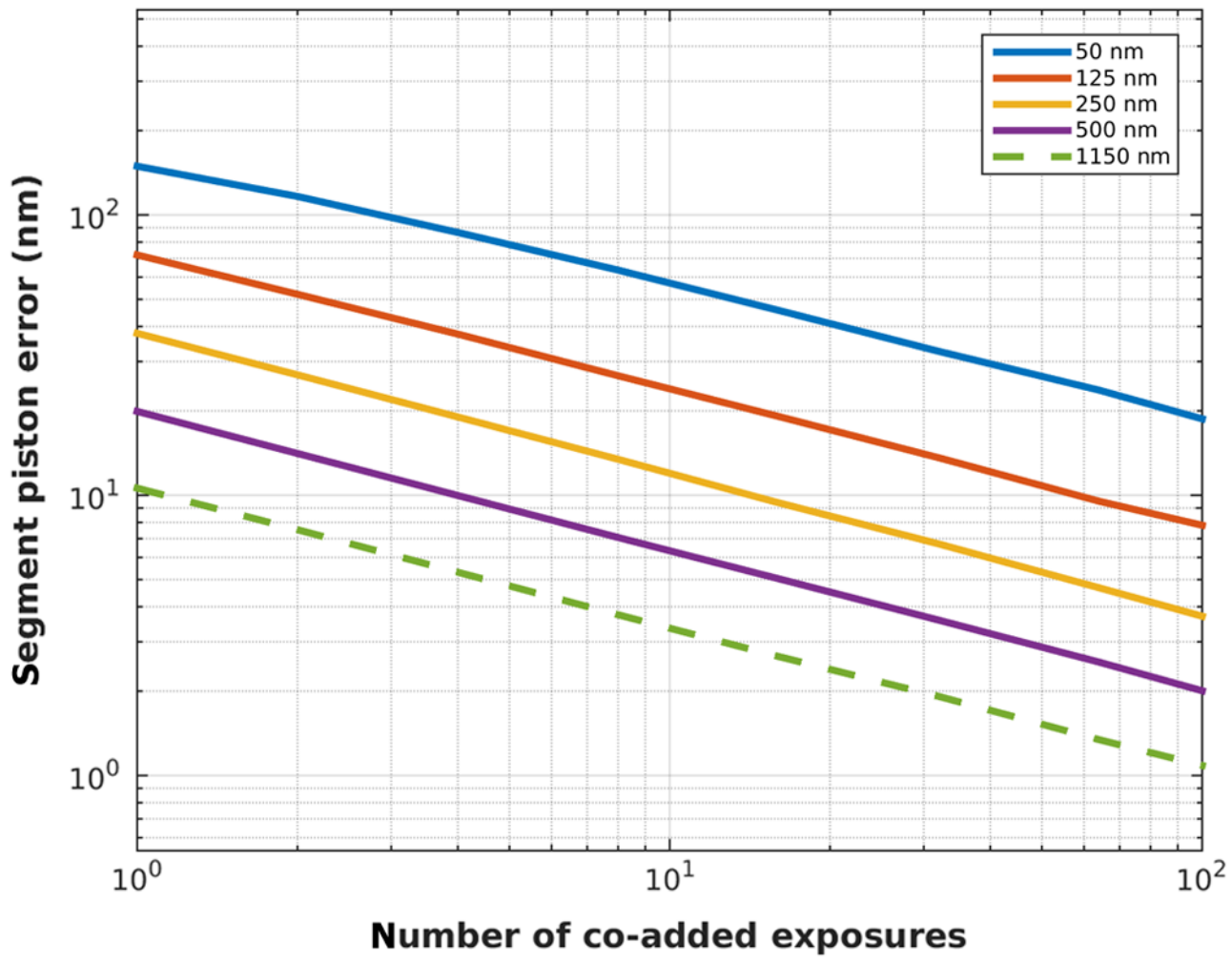

**Fig. 11** Piston estimates for various tilted segment displacement amounts versus number of co-added exposures. The dashed line is the maximum recommended segment displacement before the delta-field begins to wrap and introduce unneccesary higher spatial frequencies. Each exposure is adjusted such that the average maximum pixel value in the star image is 72,000 $e^-$. For the filter F466N, the total number of detected photons, including both the baseline and modified images, is $1.09 \times 10^6$ $e^-$ times the number of co-added exposures.

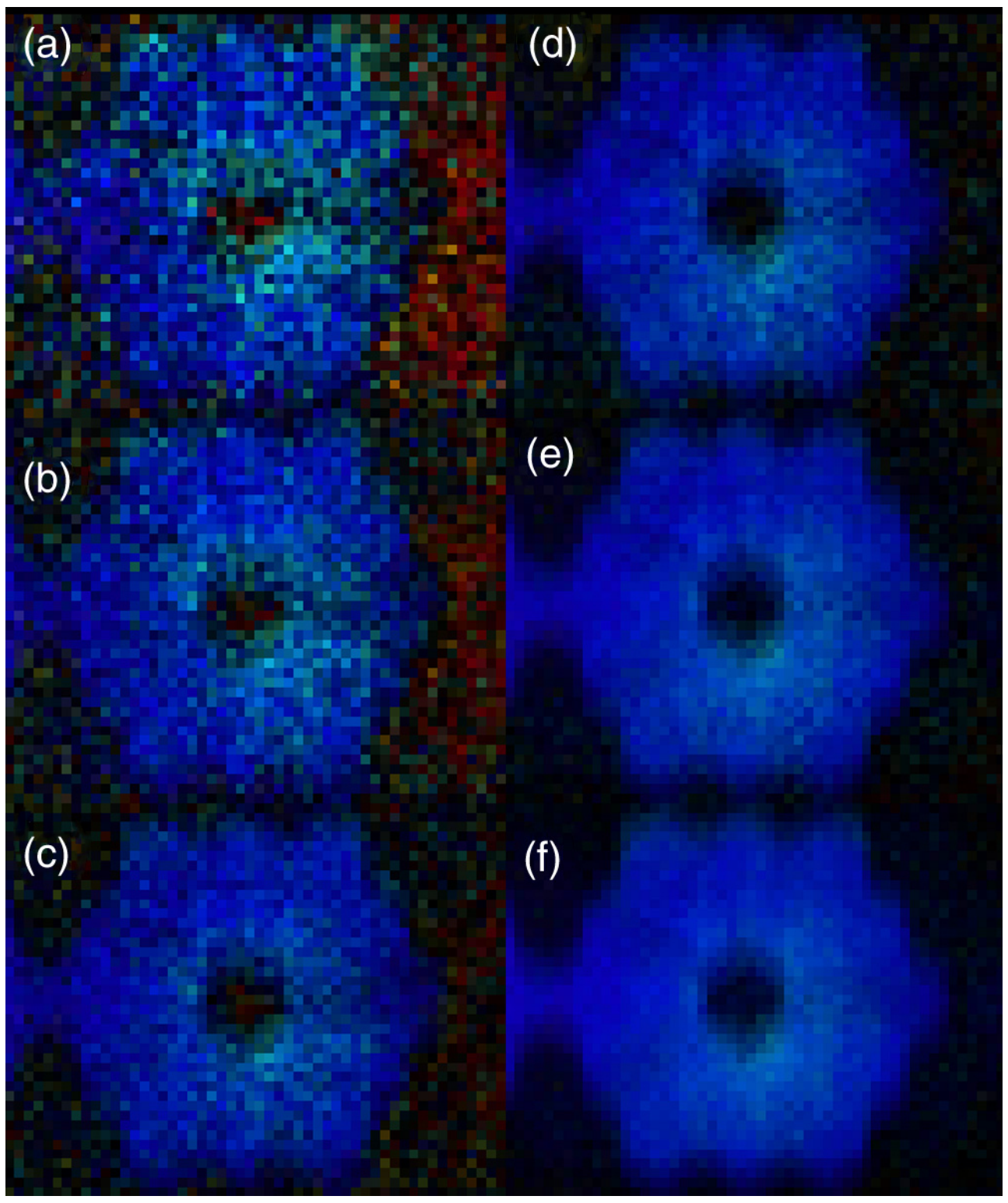

**Fig. 12** The effect of exposure time using the NIRCam filter wheel method. This example also uses the F466N filter and a 20% obscuration of the leftmost segment. The test OPD map is the same as shown in Fig. 10(a) as are the other simulation assumptions. (a) uses two single 1.85 s exposures of an eighth magnitude star. Subsequent cases show the estimated dOTF for (b) 2 exposures, (c) 4 exposures, (d) 8 exposures, (e) 16 exposures, and (f) 32 exposures.

Note that the piston accuracy averaged over an entire segment for a single pair of exposures in Fig. 11 using the $\lambda/4$ displacement is very small, $\sim 0.002\lambda$ or ∼10 nm. In practice, it will be useful to see surface height variations over regions that are much smaller than an entire segment, depending on the desired goal. The scaled accuracy from our experiment of 70 nm (Table 1) is seven times larger, but estimates the error over a smaller region and with different camera and filter details, number of detected photons, etc. The estimated accuracy quoted in the table is meant only as a rough conservative guide and can easily vary by more than a decade depending on the details of the measurement.

We also considered the effect of noise on the NIRCam filter wheel method using the same filter and parameters as with the tilted segment method. We assumed that the filter wheel edge obscured the leftmost segment by a fraction of the segment's edge-to-edge width. Examples of the resulting dOTF are shown in Fig. 12 using a single pair of images [Fig. 12(a)], two co-added images [Fig. 12(b)], four co-added images [Fig. 12(c)], eight co-added images [Fig. 12(d)], 16 co-added images [Fig. 12(e)], and 32 co-added images [Fig. 12(f)] or 64 total exposures. For our example of an 8th magnitude star and a set of 1.85 s exposures, the total exposure time runs from ∼4 s for Fig. 12(a) to ∼2 min for Fig. 12(f).

We estimated the piston error averaged over regions the size of a single segment for various amounts of pupil obscuration by the NIRCam filter wheel in Fig. 13. The obscuration amounts are listed in fractions of the edge segment's width. Note that the errors are, in general, larger than those in Fig. 11. This is mainly due to the lower dOTF signal that results when a smaller portion of a segment is involved in the pupil modification. Otherwise, the statistical characters of the signal and noise are the same.

As mentioned above, uniform noise across the individual images causes a complex Gaussian noise in the dOTF estimates which is uncorrelated even in adjacent pixels. This is very different from the complex noise that results from photon noise in the star images. The focal plane distribution of starlight photons causes this part of the dOTF noise to be spatially correlated over regions comparable with the pupil. We can see this in Fig. 14. We used the 20% blocked segment case as an example. We computed 1024 realizations and averaged the dOTF estimates to get the estimate of the mean dOTF $\langle \delta\mathcal{O} \rangle$ in Fig. 14(a). A single pair of exposures gives the dOTF realization shown in Fig. 14(b). This includes the average dOTF plus the complex noise. Since the images are real, the noise, like the dOTF, is Hermitian (symmetric real part, antisymmetric imaginary part). Figure 14(c) shows just the complex noise, given by $\delta\mathcal{O} - \langle \delta\mathcal{O} \rangle$. Notice that the noise extends beyond the average dOTF and has both small-scale and larger-scale structures. In Fig. 14(d), the small-scale noise is averaged by smoothing over a region $\sim 1/10$ the area of a single segment, leaving just the larger-scale correlated noise structure.

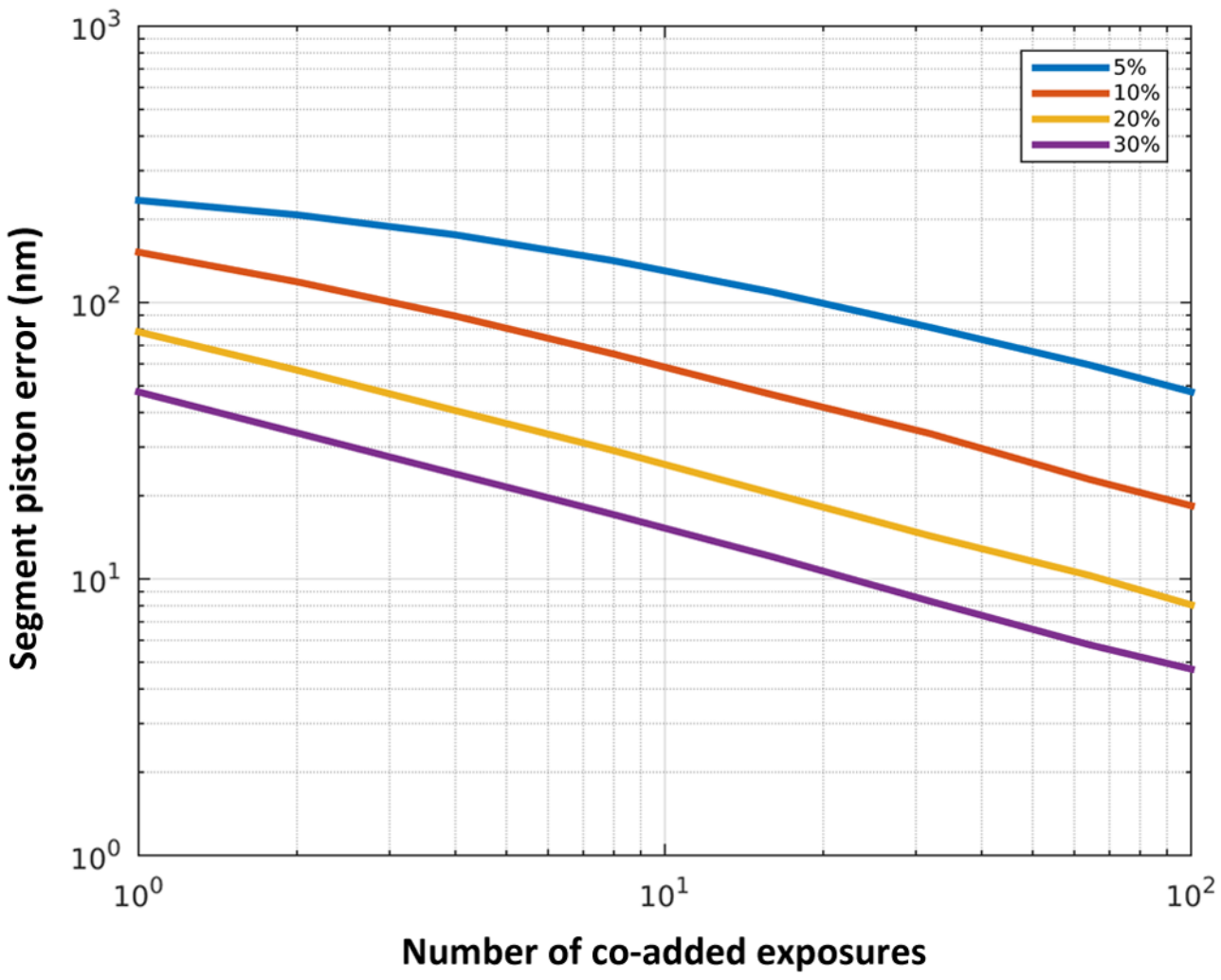

**Fig. 13** Piston estimates averaged over a segment for various amounts of pupil obscuration by the NIRCam filter wheel.

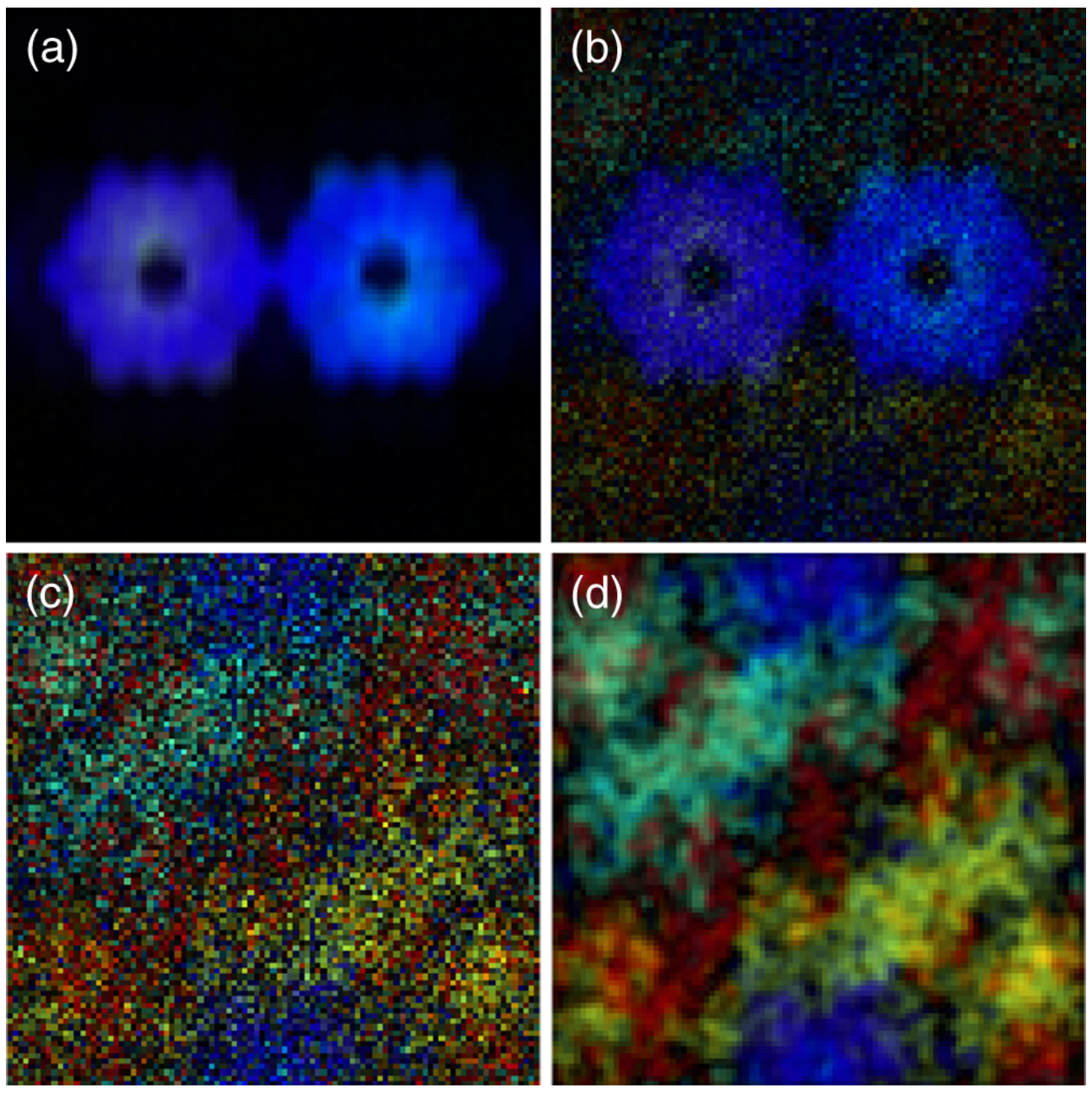

**Fig. 14** Spatial noise correlation. (a) shows the average of 1024 dOTF realizations for the filter wheel method. A single complex realization is shown in (b). Subtracting the mean (a) from the realization gives (c), which is an Hermitian noise field. The noise residual is not affected by the location of the mean dOTF pattern, but does show two different spatial scales. The background noise over the entire PSF images leads to complex noise that is uncorrelated at the dOTF pixel size. The photon noise in the star images results in spatially correlated complex noise in the dOTF, with a scale comparable with the pupil. This can be seen by applying a small amount of smoothing to the residual noise in (c), giving (d). Both the small-scale and larger-scale noise follow complex Gaussian statistics on a per-pixel basis.

## 7 Conclusions

The dOTF method is a new and simple way to measure the complex amplitude of the pupil field in an optical system, including the effects of non-common-path aberrations and transmission effects. With a straightforward theory, simulations, and laboratory experiments, we have proven that the technique works and is as simple and robust as it appears. We suggest that dOTF be considered as a supplementary or backup procedure for aligning and maintaining the segments of the JWST primary. We have identified usable filter configurations in both short- and long-wave bands for NIRCam as well as MIRI and NIRISS configurations which could be used for wavefront sensing. We presented a way to combine dOTF measurements in two filter bands to uniquely measure segment piston differences of over 0.6 mm. The dOTF measurements can be used to supplement the planned phase retrieval techniques by providing a first approximation or another constraint.

In future work, we plan to carry out a much more detailed theoretical and laboratory study of the signal and noise properties of dOTF for both the pupil blocking and phase modification methods. We will also pursue deconvolution strategies for both the spatial blurring effect and the radial blurring caused by broadband measurements. It will also be interesting to pursue other possible synergies between dOTF and phase retrieval, both in terms of computation and operational efficiency.

## Acknowledgments

This research was supported by the National Science Foundation Grants AST-0804586 and AST-0904839, NIH Grant EY020901, and NASA Grant #NNX13AC86G.

**Johanan L. Codona** is a senior research scientist with the Center for Astronomical Adaptive Optics (CAAO), Steward Observatory, University of Arizona. Until 2002, he was a Distinguished Member of Technical Staff at AT&T (later Lucent Technologies) Bell Laboratories. His interest areas include high-contrast imaging, wavefront sensing, novel uses of phase in imaging, and wave propagation through random media. He received both his PhD in applied physics and his MS in applied physics and electrical engineering from the University of California, San Diego. He received his BS in applied physics from the California Institute of Technology.

**Nathan Doble** is an associate professor in the College of Optometry at Ohio State University (OSU). His primary research interest is the development of high resolution optical imaging systems for the visualization of the human retina. He holds MSci, MSc, and PhD degrees in the fields of laser physics and adaptive optics and performed postdoctoral work at the University of Rochester, NY, USA. He was a cofounder of Iris AO Inc.